\documentclass [14pt]{article}
\usepackage[cp1251]{inputenc}
\usepackage[english,russian,ukrainian]{babel}
\usepackage{graphicx}

\begin{document}

\begin{center}
\textbf{Analytical forms of wave function and form factors of deuteron}
\end{center}

\begin{center}
\textbf{V.I. Zhaba}
\end{center}

\begin{center}
\textbf{Аналітичні форми хвильової функції та формфактори дейтрона}
\end{center}

\begin{center}
\textbf{В.І. Жаба}
\end{center}

\begin{center}
Ужгородський національний університет

вул. Волошина, 54, 88000 Ужгород, Україна
\end{center}

\textbf{Abstract}

On the received coefficients of the analytical forms for deuteron
wave function in coordinate representation in form $r^{l +
1}$\textit{*exp(-A*r)} for the modern realistic nucleon-nucleon
potentials NijmI, NijmII, Nijm93, Reid93 and Argonne v18 are
calculated charge$ G_{C}(p)$, quadrupole $G_{Q}(p)$ and magnetic
$G_{M}(p)$ deuteron form factors. An original dipole fit for
proton and neutron isoscalar electric and magnetic form factors is
used for calculations. Theoretical calculations of values of the
deuteron form factor are compared with their experimental data of
world collaborations (Bates, BLAST, Bonn, JLab, Mainz, Naval
Research Lab, NIKHEF, Orsay, Saclay, SLAC, Stanford, VEPP3 and
VEPP4) and reviews (Abbott, Boden, Garcon and Karpius). The change
of the sign of form factors is in the pulse regions at 4.7-4.9
fm$^{ - 1}$ for $G_{C}$; at 12.8-14.7 fm$^{ - 1}$ for $G_{Q}$; at
6.3-8.1 and 11.4-12.2 fm$^{ - 1}$ for $G_{M}$. The theoretical
values for the form factors $G_{Q}$(0) and $G_{M}$(0) were
calculated at the boundary condition for the momentum at
$p^{2}$=0. Calculated positions of the zero for deuteron form
factors are compared with values for other potential models. The
peculiarities of parametrizations of deuteron form factors for
theoretical approaches and fits of theoretical calculations to
experimental values, which are described in the cited literature,
are analyzed. Formulas for spherical and quadrupole form factors
are writed, which are expressed in terms of the coefficients for
the indicated deuteron wave function. At large values of momentum
the asymptotics of the deuteron form factors are determined by the
coefficients of the analytical forms for deuteron wave function,
nucleon isoscalar form factors and the order of momentum in the
denominator. The further use of deuteron form factors for
obtaining polarization observables in processes involving deuteron
as a projectile is discussed.

\textbf{Абстракт}

По отриманим раніше аналітичним формам хвильової функції дейтрона
в координатному представленні для сучасних реалістичних
нуклон-нуклонних потенціалів NijmI, NijmII, Nijm93, Reid93 і
Argonne v18 розраховано зарядовий $G_{C}$, квадрупольний $G_{Q}$ і
магнітний $G_{M}$ формфактори дейтрона. Порівнюються теоретичні
розрахунки з експериментальними даними провідних колаборацій
(Bates, BLAST, Bonn, JLab, Mainz, Naval Research Lab, NIKHEF,
Orsay, Saclay, SLAC, Stanford, VEPP3, VEPP4) та оглядів (Abbott,
Boden, Garcon, Karpius). Розраховану позицію нуля формфакторів
порівняно із значеннями для інших потенціальних моделей. При
великих значеннях імпульсу асимптотики формфакторів дейтрона
$G_{i}$ визначаються коефіцієнтами аналітичних форм хвильової
функції дейтрона, нуклонними ізоскалярними формфакторами і
степінню імпульсу в знаменнику.


\textbf{ключові слова}: дейтрон, хвильова функція, формфактор, позиція нуля, асимптотика.

\textbf{Вступ}

Дейтрон є найпростішім ядром. Простота і наочність будови дейтрона
робить його зручною лабораторією для вивчення і моделювання
нуклон-нуклонних сил. На даний час дейтрон добре вивчений
експериментально і теоретично \cite{Zhaba01, Zhaba02}.

Теоретично визначені значення статичних характеристик дейтрона
добре узгоджуються з наявними експериментальними даними. Однак
незважаючи на це, існують певні теоретичні неузгодженості і
проблеми. Наприклад, в деяких останніх роботах одна або обидві
компоненти радіальної хвильової функції дейтрона (ХФД) в
координатному представленні мають вузли поблизу початку координат.
Існування вузлів у хвильових функціях основного і єдиного стану
дейтрона свідчить про неузгодженості і неточності в реалізації
чисельних алгоритмів в розв'язанні подібних задач \cite{Zhaba01,
Zhaba02, Zhaba1}. Або це пов'язано з особливостями потенціальних
моделей для опису дейтрона.

Слід зазначити, що такі потенціали нуклон-нуклонної взаємодії, як
Боннський, Московський, потенціали Неймегенської групи
\cite{Stoks1994}, Argonne v18 \cite{Wiringa1995}, Парижський, NLO,
NNLO, N$^{3}$LO, Idaho N$^{3}$LO чи Oxford мають досить непросту
структуру і громіздку форму запис. Наприклад, оригінальний
потенціал Reid68 був параметризований на основі фазового аналізу
Неймегенською групою і отримав назву: оновлена регуляризована
версія - Reid93.

Крім того, ХФД може бути представлена таблично: через відповідні
масиви значень. При чисельних розрахунках оперувати такими
масивами чисел іноді доволі складно і взагалі незручно. І сам
текст програм для розрахунків є громіздкий, перевантажений і
нечитабельним. Тому є доцільним отримання більш простих
аналітичних форм представлення ХФД. У подальшому по них можна
розрахувати формфактори і тензорну поляризацію, що характеризують
структуру дейтрона.

ХФД у зручній формі необхідні для використання у розрахунках
поляризаційних характеристик дейтрона, а також для оцінки
теоретичних значень формфакторів дейтрона та порівняння їх з
експериментальними даними.

\textbf{Аналітичні форми хвильової функції дейтрона}

В 2000-x рр. було застосовано аналітичні ХФД в координатному
представленні. Серед них слід відмітити параметризації Дубовиченко
\cite{Dubovichenko2000}, Бережний-Корда-Гах \cite{Berezhnoy2005},
а також параметризацію у такому простому виді \cite{Zhaba1}

\begin{equation}
\label{eq1} \left\{ {\begin{array}{l}
 u(r) = r\sum\limits_{i = 1}^N {A_i e^{ - a_i r},} \\
 w(r) = r^3\sum\limits_{i = 1}^N {B_i e^{ - b_i r}.} \\
 \end{array}} \right.
\end{equation}

Незважаючи на громіздкі і довготривалі розрахунки і мінімізації
\textit{$\chi $}$^{2}$ (до величини менших за 10$^{ - 4})$,
доводилося апроксимувати за допомогою (\ref{eq1}) чисельні
значення ХФД для потенціалів Неймегенської групи (NijmI, NijmII,
Nijm93, Reid93) і потенціалу Argonne v18, масиви чисел яких
становили відповідно по 839х2 і 1500х2 значень в інтервалі
$r$=0-25 fm. Число доданків суми було вибрано $N$=11. Коефіцієнти
$A_{i}$, $a_{i}$, $B_{i}$, $b_{i}$ аналітичних форм (\ref{eq1})
для потенціалів Неймегенської групи приведені в \cite{Zhaba1}.
Розраховані ХФД не містять надлишкових вузлів поблизу початку
координат. Розраховані по ХФД статичні параметри (радіус,
електричний квадрупольний момент, магнітний момент, вклад $D$-
стану та асимптотика $D/S$- стану) і тензорна поляризація $t_{2i}$
узгоджуються з літературними теоретичними й експериментальними
даними.

В \cite{Zhaba2} проведено детальний огляд аналітичних форм для
хвильової функції дейтрона в координатному представленні.
Приведено як аналітичні форми, так і параметризації ХФД, необхідні
для подальших розрахунків характеристик процесів з участю
дейтрона.

\textbf{Формфактори дейтрона}

Для кількісного розуміння структури дейтрона, S- і D- станів та
поляризаційних характеристик розглядаються різні моделі NN
потенціалу. Розподіл заряду дейтрона добре не відомий з
експерименту, оскільки він здійснюється тільки через використання
як вимірювань поляризації, так і неполяризованих пружніх розсіяних
диференціальних перерізів. Однак його можна визначити
\cite{Abbott20001}. Диференціальний переріз пружного розсіяння
неполяризованих електронів неполяризованими дейтронами без
вимірювання поляризації відбитих електронів і дейтронів задається
формулою у рамках припущень першого Борівського наближення і умов
релятивістської інваріантності \cite{Elias1969, Galster1971,
Garcon2001, Gilman2002}

\begin{equation}
\label{eq2} \frac{d\sigma }{d\Omega _e } = \left( {\frac{d\sigma
}{d\Omega _e }} \right)_{Mott} \left[ {A(p^2) + B(p^2)tg^2\left(
{{\theta _e } \mathord{\left/ {\vphantom {{\theta _e } 2}} \right.
\kern-\nulldelimiterspace} 2} \right)} \right],
\end{equation}

де $\left( {\frac{d\sigma }{d\Omega }_e } \right)_{Mott} $ -
переріз розсіяння на безспіновій безструктурній частинці,
отриманий Моттом; \textit{$\theta $}$_{e}$ - кут розсіяння
електронів у лабораторній системі; $p$ - переданий імпульс
дейтрона в одиницях fm$^{ - 1}$; $A(p)$ і $B(p)$ - функції
електричної та магнітної структури:

\begin{equation}
\label{eq3} A = G_C^2 + \frac{8}{9}\eta ^2G_Q^2 + \frac{2}{3}\eta
G_M^2 ; \quad B = \frac{4}{3}\eta \left( {1 + \eta } \right)G_M^2
,
\end{equation}

де $\eta = \frac{p^2}{4m_d^2 }$; $m_{d}$=1875.63 МеВ - маса
дейтрона. Тут зарядовий $G_{C}(p)$, квадрупольний $G_{Q}(p)$ і
магнітний $G_{M}(p)$ формфактори містять інформацію про
електромагнітні властивості дейтрона \cite{Gilman2002, Adamu2008}:

\begin{equation}
\label{eq4} G_C = G_{EN} D_C ; \quad G_Q = G_{EN} D_Q ; \quad G_M
= \frac{m_d }{2m_p }\left( {G_{MN} D_M + G_{EN} D_E } \right),
\end{equation}

де ізоскалярний електричний і магнітний формфактори

\[
G_{EN} = G_{Ep} + G_{En} ; \quad G_{MN} = G_{Mp} + G_{Mn} .
\]

Складові формфакторів $D_{i}$ визначаються по формулам

\begin{equation}
\label{eq5} D_C = S_0^{(1)} + S_0^{(2)} ; \quad D_Q =
\frac{3}{\sqrt 2 \eta }\left( {S_2^{(1)} - \frac{1}{\sqrt 8
}S_2^{(2)} } \right);
\end{equation}

\begin{equation}
\label{eq6} D_M = 2\left( {S_0^{(1)} - \frac{1}{2}S_0^{(2)} +
\frac{1}{\sqrt 2 }S_2^{(1)} + \frac{1}{2}S_2^{(2)} } \right);
\quad D_E = \frac{3}{2}\left( {S_0^{(2)} + S_2^{(2)} } \right),
\end{equation}

де елементарні сферичні $S_0^{(i)} $ та квадрупольні $S_2^{(i)} $
формфактори \cite{Ladygin2002, Platonova20101}

\begin{equation}
\label{eq7} S_0^{(1)} = \int\limits_0^\infty {u^2j_0 dr} ; \quad
S_0^{(2)} = \int\limits_0^\infty {w^2j_0 dr} ; \quad S_2^{(1)} =
\int\limits_0^\infty {uwj_2 dr} ; \quad S_2^{(2)} =
\int\limits_0^\infty {w^2j_2 dr} .
\end{equation}

Тут $u$ і $w$ - радіальні ХФД в координатному представленні;
$j_{0}$, $j_{2}$ - сферичні функції Бесселя нульового і другого
порядку від аргументу \textit{pr}/2; $G_{Ep}$, $G_{En}$ ($G_{Mp}$,
$G_{Mn}$)- протонний і нейтронний ізоскалярний електричний
(магнітний) формфактори, які виражені через оригінальне дипольне
наближення \cite{Bekzhanov2013}. Для формфакторів $D_{i}$
виконуються умови при $p^{2}$=0 і $G_{EN} (0) = 1$; $G_{MN} (0) =
\mu _p + \mu _n $

\[
D_C (0) = 1; \quad D_Q (0) = m_d^2 Q_d ; \quad D_M (0) = 2 - 3P_D
; \quad D_E (0) = \frac{3}{2}P_D .
\]

Причому тоді самі формфактори дейтрона рівні

\[
G_C (0) = 1; \quad G_Q (0) = m_d^2 Q_d ; \quad G_M (0) = \frac{m_d
}{2m_p }\left( {(\mu _p + \mu _n )(2 - 3P_D ) + \frac{3}{2}P_D }
\right),
\]

де $P_D$, $Q_d$ - вклад $D$- стану і електричний квадрупольний
момент дейтрона:

Розраховані теоретичні значення $G_{Q}$(0) і $G_{M}$(0) для
потенціалів Неймегенської групи потенціалів (NijmI, NijmII,
Nijm93, Reid93) та потенціалу Argonne v18 (Av18) приведено в
Таблиці 1. Експериментальні значення $G_{Q}$(0) і $G_{M}$(0)
розраховані по формулам $G_Q (0) = m_d^2 Q_d $; $G_M (0) = m_d \mu
_d / m_p $.

Таблиця 1. Значення $G_{M}$(0) і $G_{Q}$(0) для різних моделей

\begin{tabular}{|l|l|l|}
\hline Потенціал& $G_{M}$(0)&
$G_{Q}$(0) \\
\hline NijmI, NijmII, Nijm93& 1.6941; 1.6943; 1.6929&
24.351; 24.214; 24.056 \\
\hline Reid93, Argonne v18& 1.6936; 1.6929&
23.947; 24.210 \\
\hline Bonn A, B, C, Q \cite{Garcon2001}& -&
24.8; 25.1; 25.4; 24.7 \\
\hline Reid SC, Paris, Av18 \cite{Garcon2001}& -&
25.2; 25.2; 24.1 \\
\hline NijmI, NijmII \cite{KhokhlovArx2016}& 1.697/1.695;
1.700/1.695&
24.8/24.6; 24.7/24.5 \\
\hline Paris, CD-Bonn \cite{KhokhlovArx2016}& 1.696/1.694;
1.708/1.704&
25.6/25.2; 24.8/24.4 \\
\hline Argonne18, JISP16 \cite{KhokhlovArx2016}& 1.696/1.694;
1.720/1.714&
24.7/24.4; 26.3/26.1 \\
\hline Moscow06, Moscow14 \cite{KhokhlovArx2016}& 1.711/1.699;
1.716/1.700&
24.5/24.2; 26.0/25.8 \\
\hline Idaho, eD \cite{KhokhlovArx2016}& 1.714/1.700; 1.715/1.700&
26.22/25.98; 25.83/25.54 \\
\hline Експеримент \cite{Garcon2001}& 1.7135&
25.833 \\
\hline
\end{tabular}

Проведено порівняння з іншими теоретичними розрахунками. Причому
дані з \cite{KhokhlovArx2016} вказані для
релятивістських/нерелятивістських результатів обчислень.

На Рис. 1-3 приведено формфактори дейтрона $G_{i}(p)$, розраховані
по ХФД (\ref{eq1}) для потенціалів NijmI, NijmII, Nijm93, Reid93,
Argonne v18 (позначення відповідно - N1, N2, N93, R93 і Av18).
Порівнюються теоретичні розрахунки з експериментальними даними:
для зарядового формфактору $G_{C}(p)$ - колаборацій Orsay
\cite{Benaksas1966}, Bates \cite{The1991, Garcon1994}, JLab
\cite{Abbott20001}, NIKHEF \cite{Bouwhuis1999}, VEPP3
\cite{Nikolenko2003, Nikolenko2010, Zevakov2006}, VEPP4
\cite{Toporkov2016}, BLAST \cite{Kohl2008, Zhang2011} та оглядів
Boden \cite{Boden1991}, Garcon \cite{Garcon1994} і Abbott
\cite{Abbott2000}; для квадрупольного формфактору $G_{Q}(p)$ -
колаборацій Orsay \cite{Benaksas1966}, Bates \cite{The1991,
Garcon1994}, JLab \cite{Abbott20001}, VEPP3 \cite{Nikolenko2003,
Nikolenko2010, Zevakov2006}, VEPP4 \cite{Toporkov2016}, BLAST
\cite{Kohl2008, Zhang2011} та оглядів Garcon \cite{Garcon1994} і
Abbott \cite{Abbott2000}; для магнітного формфактору $G_{M}(p)$ -
колаборацій Stanford \cite{Goldemberg1964, Rand1967, Rand1973},
Orsay \cite{Benaksas1966, Grossetete1966}, Naval Research Lab
\cite{Jones1980} та огляду Karpius \cite{Karpius2005}. Також
експериментальні значення магнітного формфактору отримані для
Mainz \cite{Simon1981}, Bonn \cite{Cramer1985}, Saclay
\cite{Auffret1985}, SLAC \cite{Arnold1987, Bosted1990},
використовуючи співвідношення між $G_{M}(p)$ та структурною
функцією $B$.

\pdfximage width 110mm {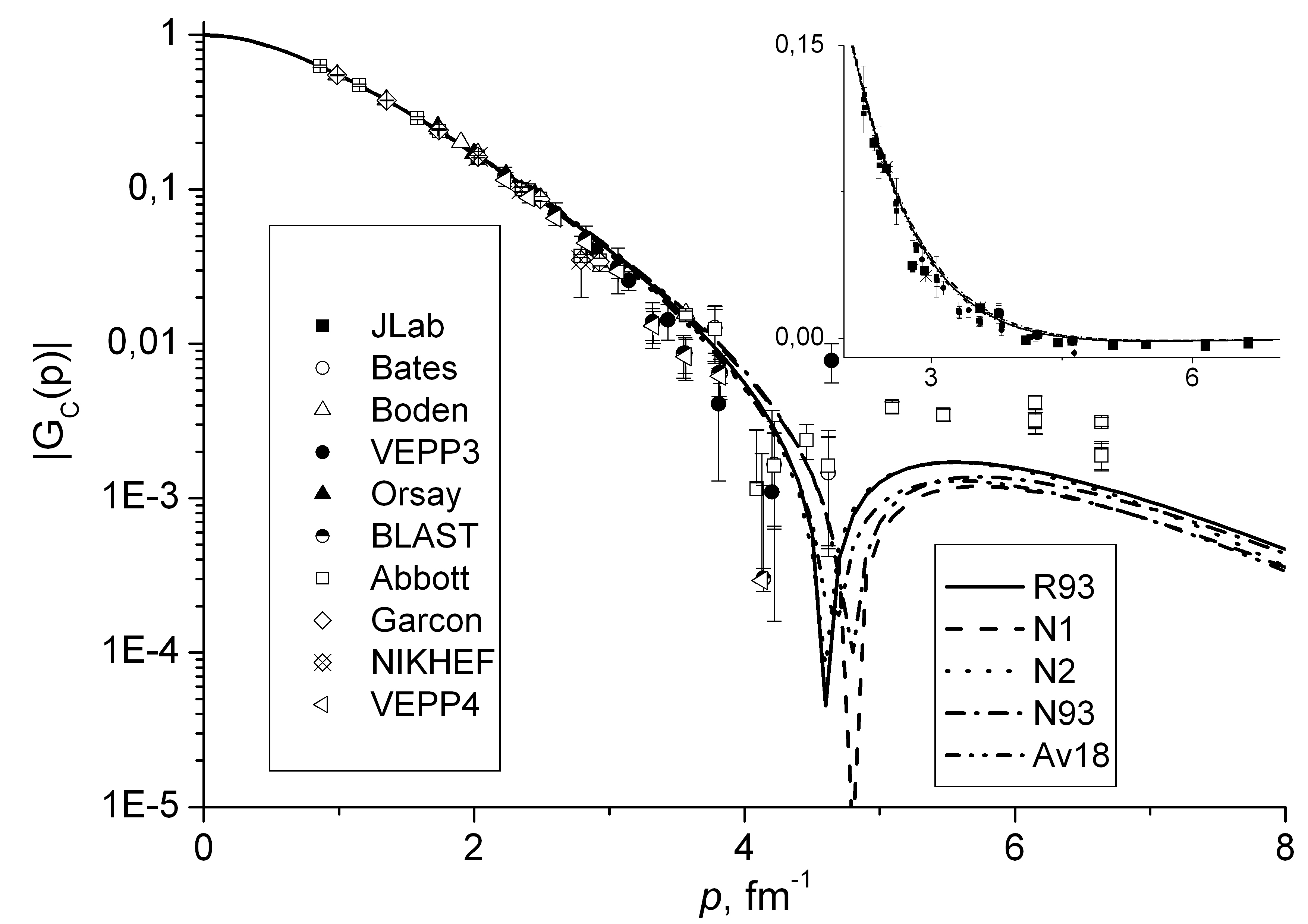}\pdfrefximage\pdflastximage

Рис.~1. Зарядовий формфактор $G_{C}(p)$.

\pdfximage width 110mm {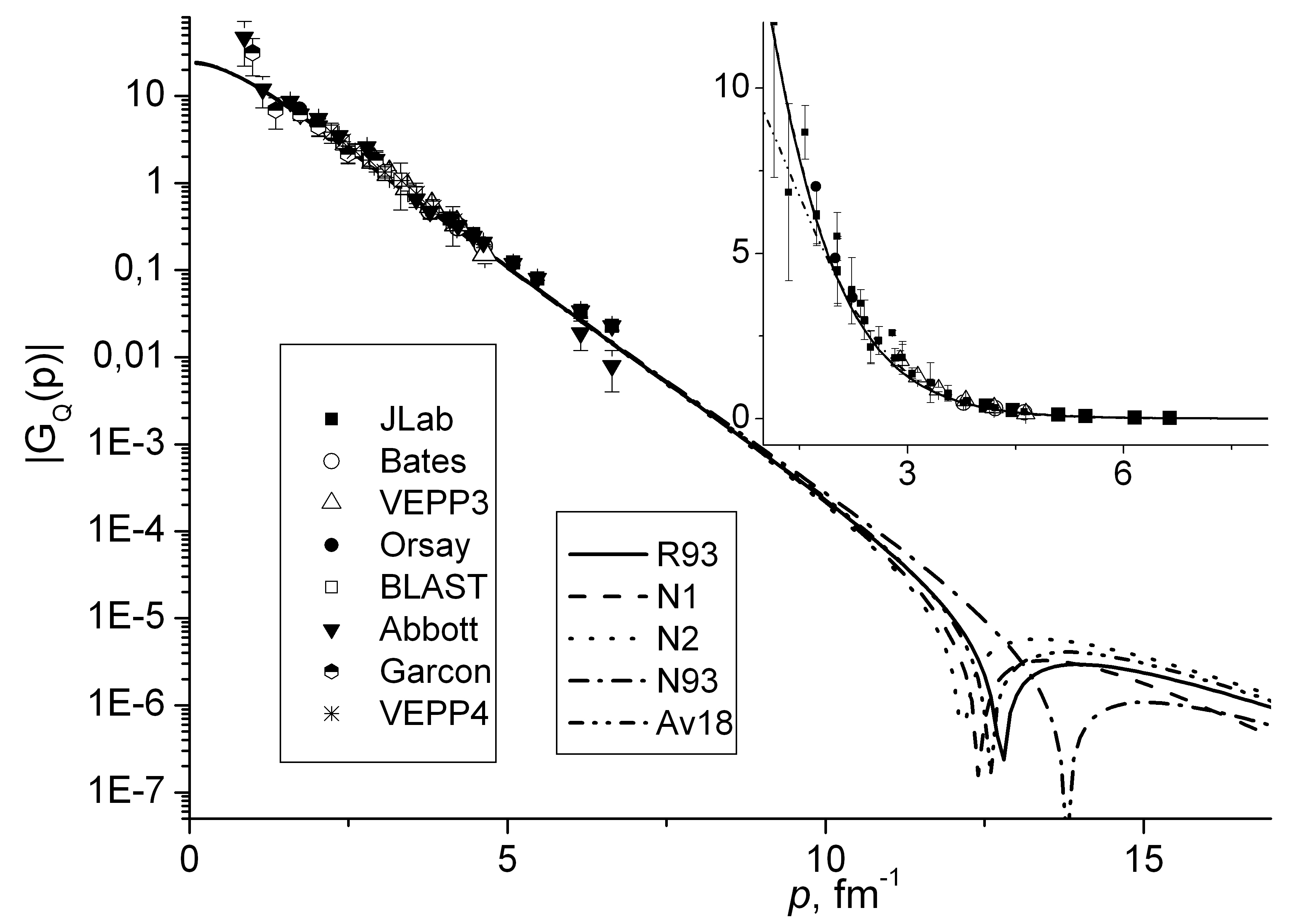}\pdfrefximage\pdflastximage

Рис.~2. Квадрупольний формфактор $G_{Q}(p)$

\pdfximage width 110mm {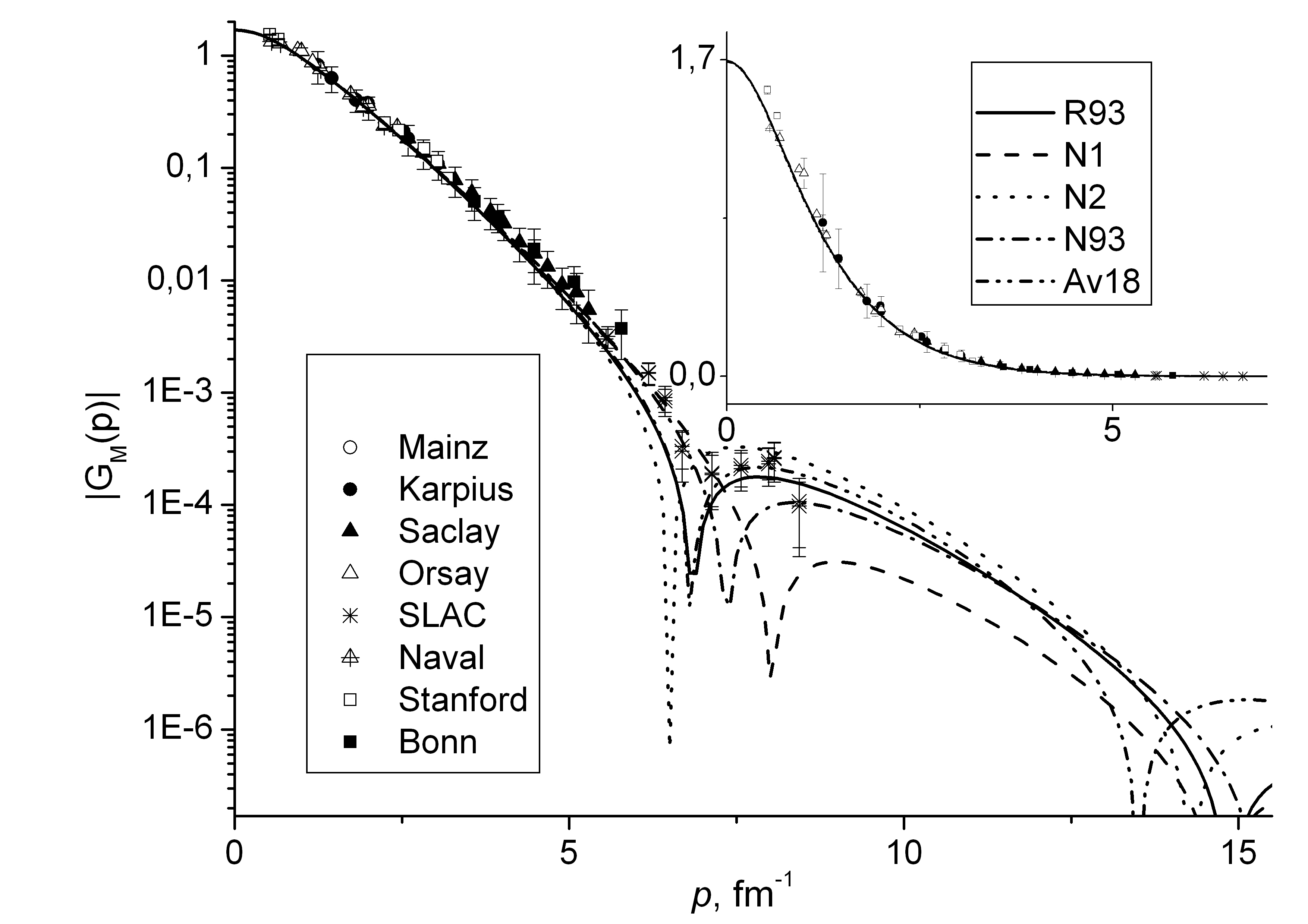}\pdfrefximage\pdflastximage

Рис.~3. Магнітний формфактор $G_{M}(p)$

Слід звернути увагу на знакозмінність формфакторів при 4.7-4.9
fm$^{-1}$ для $G_{C}(p)$; при 12.8-14.7 fm$^{-1}$ для $G_{Q}(p)$;
при 6.3-8.1 та 11.4-12.2 fm$^{-1}$ для $G_{M}(p)$.

Розрахована для потенціальних моделей позиція нуля зарядового
формфактору $G_{C}(p_{0})$ приведена в Таблиці 2. Отримані
значення порівняно з даними для інших потенціальних моделей та з
експериментальним значенням. Позначення в Таблицях приведено
згідно цитованої літератури.

Таблиця 2. Позиція нуля зарядового формфактору $G_{C}$

\begin{tabular}{|l|l|}
\hline Потенціал&
$p_{0}$, fm$^{-1}$ \\
\hline NijmI, NijmII, Nijm93&
4.80;4.60; 4.81 \\
\hline Reid93, Argonne v18&
4.60; 4.71 \\
\hline Bonn-B, FULLF, OBEPF \cite{Plessas1995}&
4.28; 4.26; 4.10 \\
\hline 5RM \cite{Gilman2002}, 4RIA \cite{Garcon1994}&
4.8-5.1; 4.3-4.6 \\
\hline Moscow, NijmI, NijmII \cite{Khokhlov2015}&
4.1; 4.6; 4.4 \\
\hline CD-Bonn, Paris \cite{Khokhlov2015}&
4.4; 4.9 \\
\hline TР \cite{Kohl2008}, NijmI, NijmII \cite{Krutov2002}&
4.19; 4.9; 4.7 \\
\hline CIA, RIA \cite{VanOrden1995}&
4.24-4.53 \\
\hline CD-Bonn, Paris, РІН \cite{Krutov2002}, NIA
\cite{Adamu2008}&
5.2; 4.6; 4.3; 4.6 \\
\hline 9RIA \cite{Arnold1980}&
4.5; 9.2-12.3 \\
\hline 9DWF \cite{Mathelitsch1978}, 5NC \cite{Bouwhuis1999}&
4.5-6.0; 3.8-4.7 \\
\hline IA AV14, IA+MEC AV14 \cite{The1991}&
4.5; 4.0 \\
\hline Paris, Av14, Bonn-E \cite{Garcon1994}&
4.5; 4.4; 5.3 \\
\hline TCM \cite{Tomasigustafsson2006}, QCB82, QCB86
\cite{Dijk1989}&
4.2; 4.4; 4.5 \\
\hline MMQCM \cite{Ito1987}, Graz-II \cite{Bekzhanov2013}&
4.5-5.2; 5.67 \\
\hline RIA \cite{Braun1991}&
4.5-5.2; 11.4-13 \\
\hline QCM \cite{Faessler1993}; OBEPQ-A, B, C
\cite{Arenhovel2000}&
4.4; 4.1-4.5 \\
\hline Експеримент \cite{Garcon1994}&
4.30$\pm $0.16 \\
\hline
\end{tabular}

Позиції нуля для квадрупольного і магнітного формфакторів (Таблиці
3 і 4) знаходяться значно правіше, ніж для зарядового формфактору.
На жаль, відсутні експериментальні значення для імпульсів більше 7
і 9 fm$^{ - 1}$ для формфакторів $G_{Q}$ і $G_{M}$ відповідно.

Таблиця 3. Позиція нуля квадрупольного формфактору $G_{Q}$

\begin{tabular}{|l|l|}
\hline Потенціал&
$p_{0}$, fm$^{-1}$ \\
\hline NijmI, NijmII, Nijm93&
12.4; 12.2; 13.8 \\
\hline Reid93, Argonne v18&
12.8; 12.6 \\
\hline 9RIA \cite{Arnold1980}&
8.7-10.9 \\
\hline RIA \cite{Braun1991}&
8.9-12.2 \\
\hline
\end{tabular}

Таблиця 4. Позиція нуля магнітного формфактору $G_{M}$

\begin{tabular}{|l|l|}
\hline Потенціал&
$p_{0}$, fm$^{-1}$ \\
\hline NijmI&
8.0; 14.7 \\
\hline NijmII&
6.5; 14.3 \\
\hline Nijm93&
7.4; 15.3 \\
\hline Reid93&
6.9; 14.9 \\
\hline Argonne v18&
6.8; 13.5 \\
\hline Moscow, NijmI, NijmII \cite{Khokhlov2015}&
5.3; 6.9; 6.0 \\
\hline CD-Bonn, Paris \cite{Khokhlov2015}&
6.7; 5.8 \\
\hline 9DWF \cite{Mathelitsch1978}, TCM
\cite{Tomasigustafsson2006}&
5.5-7.5; 7.2 \\
\hline QCB82, QCB86 \cite{Dijk1989}, QCM \cite{Faessler1993}, NIA
\cite{Adamu2008}&
7.1; 7.2; 6.5; 6.6 \\
\hline 5RM \cite{Gilman2002}, MMQCM \cite{Ito1987}&
5.6-6.1; 6.8-7.5 \\
\hline RIA \cite{Braun1991}&
6.5; 11.1 \\
\hline
\end{tabular}

\textbf{Параметризація формфакторів дейтрона}

У роботі \cite{Gilman2002} використовуються функції підгонки
теоретичних розрахунків до експериментальних значень формфакторів
дейтрона

\[
G_C^{(fit)} = e^{ - Q^2 / 3.5}F; \quad G_Q^{(fit)} =
\frac{25.8298}{1.01}\left( {e^{ - Q^2} + 0.01e^{ - Q^2 / 100}}
\right)F; \quad G_M^{(fit)} = 1.7487e^{ - Q^2 / 2.5}F;
\]

де $F = \left( {1 + \frac{Q^2}{m\varepsilon }} \right)^{ -
1}\left( {1 + \frac{Q^2}{0.71}} \right)^{ - 2}$; $Q^{2}$ - квадрат
імпульсу в GeV$^{2}$; \textit{m$\varepsilon $}=0.05365 GeV$^{2}$.
В \cite{Gilman2002} розрахунки формфакторів дейтрона по цим
формулам дають задовільні значення відносно експериментальних
даних тільки в області 0-0.5 GeV.

В огляді \cite{Abbott2000} наявні ще дві форми параметризації

\[
G_i (p^2) = G_i (0)\left[ {1 - \left( {\frac{p}{p_i^0 }}
\right)^2} \right]\left[ {1 + \sum\limits_{k = 1}^5 {a_{ik}
p^{2k}} } \right]^{ - 1}; \quad \left( {\begin{array}{l}
 G_C \\
 G_Q \\
 G_M \\
 \end{array}} \right) = G_D^2 \left( {\frac{p^2}{4}} \right) \cdot M(\eta
)\left( {\begin{array}{l}
 g_0 \\
 g_1 \\
 g_2 \\
 \end{array}} \right);
\]

де амплітуди параметризовані у виді лоренціану $g_k =
p^k\sum\limits_{i = 1}^4 {\frac{a_{ki} }{\alpha _{ki} + p^2}} $.
Для 18 і 12 параметрів для кожного формфактора ці параметризації
дали $\chi ^{2}$/$N$=1.5 і 1.8 відповідно, де $N$ - кількість
експериментальних даних. Крім цього, параметризація формфакторів
дейтрона у виді добутку гаусіана на суму (SOG) \cite{Sick1974}
була застосована для експериментальних даних огляду
\cite{Abbott2000}. Для 33 параметрів цієї параметризації $\chi
^{2}$/$N$=1.5.

В нерелятивістському імпульсному наближенні ведучі доданки
формфакторів дейтрона визначаються коефіцієнтами розкладу
($C_{j}$, $m_{j})$ в координатному представлені та степінню
імпульсу в знаменнику \cite{Krutov2005}

\begin{equation}
\label{eq8} G_C^{NR} \sim \mbox{ - }\frac{1}{p^3}\left( {G_{Ep} +
G_{En} } \right)f(C_j ,m_j ); \quad G_Q^{NR} \sim \mbox{ -
}\frac{m_d^2 }{p^7}\left( {G_{Ep} + G_{En} } \right)f(C_j ,m_j );
\end{equation}

\begin{equation}
\label{eq9} G_M^{NR} \sim \mbox{ - }\frac{1}{p^3}\left( {G_{Mp} +
G_{Mn} } \right)f(C_j ,m_j );
\end{equation}

тобто головні члени розкладу зарядового і магнітного формфакторів
дейтрона визначаються тільки S- станом хвильової функції дейтрона,
а D- компонента дає додатковий вклад в головний доданок
квадрупольного формфактору. Швидке спадання електричного
нуклонного формфактору в порівнянні із магнітним призводить до
швидшого спадання зарядового і квадрупольного формфакторів, ніж
магнітного $G_{M}$. В релятивістському імпульсному наближенні
формфактори дейтрона визначаються вищевказаним нерелятивістським
набором формфакторів:

\[
G_i^R \sim \sqrt {\frac{m_d }{p}} G_i^{NR} ;i = \left\{ {C,Q,M}
\right\}.
\]

Для ХФД (\ref{eq1}) сферичні та квадрупольні формфактори
(\ref{eq7}) будуть записані у формі

\begin{equation}
\label{eq10} S_0^{(1)} = 32\sum\limits_{i,j = 1}^N {\frac{A_i A_j
a_{ij} }{\left( {p^2 + 4a_{ij}^2 } \right)^2}} ; \quad S_0^{(2)} =
61440\sum\limits_{i,j = 1}^N {\frac{B_i B_j b_{ij} \left( {3p^4 -
40b_{ij}^2 p^2 + 48b_{ij}^4 } \right)}{\left( {p^2 + 4b_{ij}^2 }
\right)^6}} ;
\end{equation}

\begin{equation}
\label{eq11} S_2^{(1)} = - 3072\sum\limits_{i = 1}^N {\frac{A_i B_i
(a_i + b_i )p^2}{\left( {p^2 + 4(a_i + b_i )^2} \right)^4}} ;
\quad S_2^{(2)} = - 98304\sum\limits_{i,j = 1}^N {\frac{B_i B_j
b_{ij} p^2\left( {3p^2 - 28b_{ij}^2 } \right)}{\left( {p^2 +
4b_{ij}^2 } \right)^6}} ;
\end{equation}

де $a_{ij} = a_i + a_j $; $b_{ij} = b_i + b_j $. Тоді, враховуючи
складові формфакторів $D_{i}$ згідно (\ref{eq5}) та (\ref{eq6}),
самі формфактори дейтрона $G_{i}$ (\ref{eq4}) матимуть наступні
асимптотики при великих значеннях імпульсу

\begin{equation}
\label{eq12} G_C \approx 32G_{EN} \sum\limits_{i,j = 1}^N
{\frac{A_i A_j a_{ij} }{\left( {p^2 + 4a_{ij}^2 } \right)^2}} ;
\quad G_Q \approx - \frac{9216G_{EN} }{\sqrt 2 \eta
}\sum\limits_{i = 1}^N {\frac{A_i B_i (a_i + b_i )p^2}{\left( {p^2
+ 4(a_i + b_i )^2} \right)^4}} ;
\end{equation}

\begin{equation}
\label{eq13} G_M \approx \frac{m_d }{m_p }G_{MN} \left(
{32\sum\limits_{i,j = 1}^N {\frac{A_i A_j a_{ij} }{\left( {p^2 +
4a_{ij}^2 } \right)^2}} - \frac{3072}{\sqrt 2 }\sum\limits_{i =
1}^N {\frac{A_i B_i (a_i + b_i )p^2}{\left( {p^2 + 4(a_i + b_i
)^2} \right)^4}} } \right).
\end{equation}

І, підставивши величними $G_{EN}$, $G_{MN}$ і $\eta $, можна
записати асимптотики формфакторів дейтрона $G_{i}$ для великих
значень імпульсу у виді:

\begin{equation}
\label{eq14} G_C \sim \frac{1}{p^8}; \quad G_Q \sim
\frac{1}{p^{12}}; \quad G_M \sim \frac{1}{p^8}.
\end{equation}

Отже, асимптотики формфакторів дейтрона визначаються коефіцієнтами
аналітичних форм ХФД, нуклонними ізоскалярними електричним та
магнітним формфакторами та степінню імпульсу в знаменнику. Крім
цього, знаючи асимптотики формфакторів дейтрона, можна визначити
їх вклад в структурні функції.

\textbf{Висновки}

Використовуючи відомі аналітичні форми ХФД в координатному
представленні для потенціалів NijmI, NijmII, Nijm93, Reid93,
Argonne v18, розраховано зарядовий $G_{C}$, квадрупольний $G_{Q}$
і магнітний $G_{M}$ формфактори дейтрона. Порівнюються теоретичні
розрахунки з експериментальними даними провідних колаборацій
(Bates, BLAST, JLab, Naval Research Lab, NIKHEF, Orsay, Stanford,
VEPP3) та оглядів (Abbott, Boden, Garcon). Розраховану позицію
нуля формфакторів порівняно з даними для інших потенціальних
моделей.

При великих значеннях імпульсу асимптотики формфакторів дейтрона
$G_{i}$ (\ref{eq12})-(\ref{eq14}) визначаються коефіцієнтами
аналітичних форм ХФД, нуклонними ізоскалярними формфакторами, а
також степінню імпульсу в знаменнику.

Знаючи формфактори дейтрона, можна визначити такі величини як
функції електричної та магнітної структури $A(p)$ і $B(p)$,
тензорні $t_{2j}$ і векторні $t_{1i}$ поляризації
\cite{Gilman2002}, тензорну аналізуючу здатність $A_{yy}$ і
векторні (тензорні) коефіцієнти передачі поляризації $k_a^{a'} $
($k_{aa}^{a'a'} )$ в рамках моделі обміну $\omega $-мезоном
\cite{Rekalo1996}, коефіцієнти спінової кореляції $C_{xz}^{(0)} $,
$C_{zz}^{(0)} $ та тензорні асиметрії $A_{xx}^{(0)} $,
$A_{xz}^{(0)} $, $A_{zz}^{(0)} $ для пружного лептон-дейтронного
розсіяння в межах нульової лептонної маси \cite{Gakh2014} та інші
поляризаційні спостережувані в процесах за участю дейтрона.

Отримані результати дозволяють в подальшому вивчити
електромагнітну структуру дейтрона, диференціальний переріз
подвійного розсіяння, а також обчислювати теоретичні значення
зарядового і магнітного радіусів дейтрона \cite{Arenhovel2000} та
спінових спостережуваних \textit{dp}-розсіяння.

\end{document}